# DCASE 2018 CHALLENGE: SOLUTION FOR TASK 5

*Jeremy Chew*, *Yingxiang Sun*, *Lahiru Jayasinghe*, *Chau Yuen*

Singapore University of Technology and Design, Singapore

jeremycwj@gmail.com, yingxiang_sun@mymail.sutd.edu.sg, {aruna_jayasinghe, yuenchau}@sutd.edu.sg

## ABSTRACT

To address Task 5 in the Detection and Classification of Acoustic Scenes and Events (DCASE) 2018 challenge, in this paper, we propose an ensemble learning system. The proposed system consists of three different models, based on convolutional neural network and long short memory recurrent neural network. With extracted features such as spectrogram and mel-frequency cepstrum coefficients from different channels, the proposed system can classify different domestic activities effectively. Experimental results obtained from the provided development dataset show that good performance with F1-score of 92.19% can be achieved. Compared with the baseline system, our proposed system significantly improves the performance of F1-score by 7.69%.

## 1. INTRODUCTION

Intelligent housing system has attracted much attention in recent years due to people's increasing need for quality of living requirements. It is especially helpful to address the aging problem which our society is facing, since a large proportion of elderly people live independently in their own homes and monitoring their activities is imperative [1]. Various types of sensors [2]-[5] can be utilized for this monitoring purpose, among which microphone is widely used because of its cost-effectiveness. By collecting acoustic signals, not only activities of daily living [6] can be studied, but also emergency situations such as falls [7] can be detected and alerted timely.

The Detection and Classification of Acoustic Scenes and Events (DCASE) challenge is an official IEEE Audio and Acoustic Signal Processing (AASP) challenge. It provides a publicly available dataset with the participants to develop computational scene and event analysis methods. DCASE 2018 consists of five different tasks and multi-channel dataset are offered compared to the previous editions [8]. In DSASE 2018, Task 5 named "monitoring of domestic activities based on multi-channel acoustics" is highly related to the monitoring purpose in intelligent housing systems.

This paper presents a new neural network based ensemble learning system which can effectively address Task 5. The results of the proposed system and the comparison to the given baseline system are provided in detail.

## 2. PROBLEM SETUP

The main target of this study is to classify multi-channel audio segments or, acoustic scenes into some provided predefined classes. The dataset is collected using the daily activities of a person, performed in a home environment. Moreover, there are no overlapping activities exist in the dataset.

### 2.1. Dataset

The dataset used in this task is a derivative of the SINS dataset [9]. The SINS dataset also contains a one person living in a home over a period of one week. The data collected using a network of 13 microphone arrays distributed over the home and one such microphone array consists of four linearly arranged microphones. More information about the SINS dataset can be found in [10]. The dataset consists of development and evaluation data separately. The evaluation set is 1/3 of the total data and rest belongs to the development set. Since they were continuing recordings, ten seconds of splits of audio segments were generated. All the audio segment contains four channels which represent four microphone channels from a node. Segments which contain overlapping classes were omitted. Therefore, the development set is labeled in nine categories as shown in Table I. The floor plane of the home is shown as in Fig.1.

Table I  Available segments and session in development set

| Set | Activity | Segments | Sessions |
|---|---|---|---|
| Development | Absence | 18860 | 42 |
| | Cooking | 5124 | 13 |
| | Dishwashing | 1424 | 10 |
| | Eating | 2308 | 13 |
| | Other | 2060 | 118 |
| | Social Activity | 4944 | 21 |
| | Vacuum Cleaning | 972 | 9 |
| | Watching TV | 18648 | 9 |
| | Working | 18644 | 33 |

Given the dataset is imbalanced, it consists of recordings when a person is present in another room and it is noticeable in the recordings also. The "Other" class is referred to being present in the room but not doing any other activities mentioned in the Table 1.

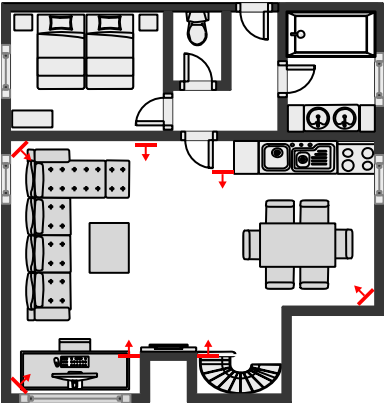

Fig.1  Floor plane of the home

### 2.2. Baseline system

The given baseline system is the benchmark for the performance analysis. The system has all the functionality for dataset handling, calculating features and models, and evaluating the results. The baseline system employed a single classifier model that takes a single channel as input. Its is Neural Network architecture with two convolutional layers and one dense layer. Log mel-band energies are provided as inputs to the network. Features are extracted for each microphone channel and at the end single outcome was generated. The baseline method used 1D convolution over the time axis and finally a fully connected layer followed by a softmax classifier will classify the input data into the given nine classes. Refer to [11] for more details about the baseline system.

### 2.3. Convolutional Neural Networks

Inspired by the breakthrough findings of Fukushima's hierarchical model called Neocognitron [12]. Thereafter, first CNN was proposed by LeCun et al. [13]. The CNN has equipped with the backpropagation (BP) algorithm for learning a receptive field. A CNN is a multilayer perceptron and mainly consists of convolutional layers and sub-sampling layers [14].

Let $d^{(l-1)}$ and $d^{(l)}$ be the input and output of the $l^{th}$ layer respectively. Input to $l^{th}$ layer is the output of $(l-1)^{th}$ layer. Since there are several number of feature maps for a layer $l$, one such feature map can be considered as $d_j^{(l)}$. Then the output of a convolutional layer is computed as

$$d_j^{(l)} = f\left(\sum_i d_i^{(l-1)} * w_{ij}^{(l)} + b_j^{(l)} 1_{s_o^{(l)}}\right) \quad (1)$$

where $*$ denotes the convolution operation, $s_o$ denotes the output dimensions and $w_{ij}^{(l)}, b_j^l$ denote the kernel weights of the $l^{th}$ layer. The $f$ is a non-linear activation function and often it would be ReLu.

The subsampling or pooling layer used to reduce the feature map resolution to increase the invariance of features to the distortions on the inputs.

$$d_j^{l+1} = maxpooling(d_j^{(l)}) \quad (2)$$

According to Eq. 2, the max-pooling output will propagate to the next layer as the input.

### 2.4. Long-Short Term Memory networks (LSTM)

A given input sequence $x = (x_1, ..., x_T)$, Recurrent Neural Networks (RNN) generate outputs of $y = (y_1, ..., y_T)$ using the hidden vector sequence of $h = (h_1, ..., h_T)$ by iterating the following equations from t=1 to T:

$$h_t = f(W_{xh}x_t + W_{hh}h_{t-1} + b_h) \quad (3)$$

$$y_t = W_{hy}h_t + b_y \quad (4)$$

where the W and b terms denote the weight matrices between layers and $f$ denotes the activation function of the layer. However, calculation of $h_t$ for LSTM networks is different

$$i_t = \sigma(W_{xi}x_t + W_{hi}h_{t-1} + W_{ci}c_{t-1} + b_i) \quad (4)$$

$$f_t = \sigma(W_{xf}x_t + W_{hf}h_{t-1} + W_{cf}c_{t-1} + b_f) \quad (5)$$

$$c_t = f_t c_{t-1} + i_t \tanh(W_{xc}x_t + W_{hc}h_{t-1} + b_c) \quad (6)$$

$$o_t = \sigma(W_{xo}x_t + W_{ho}h_{t-1} + W_{co}c_t + b_o) \quad (7)$$

$$h_t = o_t \tanh(c_t) \quad (8)$$

where σ is the logistic sigmoid function, and i, f, o and c are respectively the input gate, forget gate, output gate and cell activation vectors. Since there are several gates, this cell can keep a selective memory compared to RNN cells [15].

## 3. SYSTEM MODEL

### 3.1. Signal Pre-processing

The acoustic signals will be pre-processed before extracting features from them. Two different features are extracted, i.e., grayscale image of power spectrogram and mel-frequency cepstrum coefficients (MFCCs).

For the power spectrogram extraction, firstly, the acoustic signals are framed by a window with frame length of $L$ and hop length of $H$. Next, short-time Fourier transform (STFT) is carried out to obtain the magnitude spectrum of each frame in frequency domain. Then, the power spectrum is obtained by squaring the magnitude, which can be finally converted in to decibel units by taking logarithm.

For the MFCCs extraction, once the power spectrogram is computed, the mel filterbank will be applied to it and energy in each filter will be summed up. By taking the logarithm and the discrete cosine transform (DCT) of all filterbank energies, the DCT coefficients will be obtained. The MFCCs feature if finally obtained by only keeping the first $M$ DCT coefficients while discarding the rest.

### 3.2. System Architecture

The overall system diagram is depicted in Fig.2. As the power spectrogram image is used as one feature, we input it into 2D CNN to classify. For the MFCC feature, we use 1D CNN and LSTM as the classifier. During training procedure, the dataset is shuffled, which results in the time-independence among different batches. Therefore, we use stateless LSTM. Once all models are already trained, the models are then combined as a system, by weighted averaging output probabilities from each model. The weightages are denoted as $w_1$, $w_2$, and $w_3$, corresponding to 2D CNN, 1D CNN, and stateless LSTM. The architectures of the 2D CNN, 1D CNN, and stateless LSTM are shown as in Tables II(a), (c), and (c) respectively.

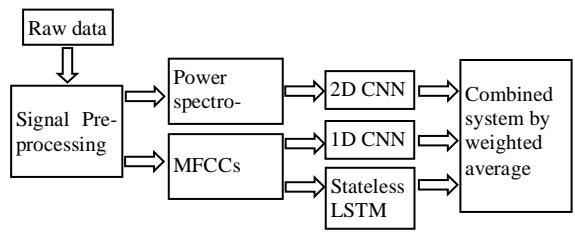

Fig.2 The overall system diagram

Table II (a) The architecture of 2D CNN

| Layer Name | Apply |
|---|---|
| Input | Power spectrogram image with size 64×64 |
| 2D Conv1 | Convolution layer: 32 filters, kernel size 3×3, strides 1×1, ReLU activation. Max pooling layer: kernel size 2×2. |
| 2D Conv2 | Convolution layer: 64 filters, kernel size 3×3, strides 1×1, ReLU activation. Max pooling layer: kernel size 2×2. |
| 2D Conv3 | Convolution layer: 64 filters, kernel size 3×3, strides 1×1, ReLU activation. Max pooling layer: kernel size 2×2. |
| Flatten | |
| Dense | 128 units, dropout rate: 0.5 |
| Softmax | 9 classes |

Table II (b) The architecture of 1D CNN

| Layer Name | Apply |
|---|---|
| Input | MFCCs |
| 1D Conv1 | Convolution layer: 128 filters, kernel size 100, strides 1, ReLU activation. |
| 1D Conv2 | Convolution layer: 128 filters, kernel size 30, strides 1, ReLU activation. Max pooling layers using kernels size 2. |
| 1D Conv3 | Convolution layer: 128 filters, kernel size 15, strides 1, ReLU activation. Max pooling layer: kernel size 2. |
| Flatten | |
| Dense | 128 units, dropout rate: 0.5 |
| Softmax | 9 classes |

Table II (c) The architecture of stateless LSTM

| Layer Name | Apply |
|---|---|
| Input | MFCCs |
| LSTM1 | 64 units, return_sequence: Ture |
| LSTM2 | 64 units, return_sequence: Ture |
| LSTM3 | 64 units |
| Dropout | dropout rate: 0.5 |
| Softmax | 9 classes |

## 4. EXPERIMENTAL RESULTS AND DISCUSSION

To develop our model, we split the given development dataset into three parts. 75% of the dataset is used as training data; 5% of the dataset is used as validation data; the rest 20% is used as testing data. In the development dataset, there are nine different domestic activities in total, including absence (AB), cooking (CO), Dishwashing (DW), Eating (EA), other (OT), social activity (SA), vacuum cleaning (VC), watching TV (WT), and working (WO).

The frame length and hop length are chosen to be 2048 and 512 respectively, while the size of grayscale image is selected as 64×64. The MFCCs feature keeps the first 20 MFCCs. We choose Adam optimizer with learning rate 1e-5 and train each model 500 epochs with batch size of 50. $w_1$, $w_2$, and $w_3$ are selected as 0.3, 0.4, and 0.3 respectively. The macro-averaged results of nine different activities obtained by each model from the testing data is shown as in Table III.

Table III Results obtained from the testing data

| Model | Macro-averaged Accuracy | Macro-averaged F1-score |
|---|---|---|
| 2D CNN | 86.90% | 86.55% |
| 1D CNN | 89.69% | 89.59% |
| Stateless LSTM | 88.07% | 88.01% |

From the macro-averaged results, we can observe that the 1D CNN model preforms best among all three models, achieving accuracy of 88.07% while F1-score of 88.01%.

The performances of 2D CNN model and stateless LSTM model are a bit lower than that of 1D CNN model. Compared to the baseline system, whose macro-averaged F1-score is 84.50%, the 2D CNN model, 1D CNN model, and stateless LSTM model improve the performance by 2.05%, 5.09%, and 3.51% respectively.

For the results of the combined system, the confusion matrix is shown as in Table IV. All the activities of watching TV are correctly classified, while those of cooking, dishwashing, eating, and social activity are also easy to distinguish. In contrast, the activities of absence, other, and working are much more difficult to discriminate.

Table IV Confusion matrix of the combined system

|    | AB  | CO  | DW  | EA  | OT  | SA  | VC  | WT  | WO  |
|----|-----|-----|-----|-----|-----|-----|-----|-----|-----|
| AB | 201 | 0   | 0   | 0   | 7   | 1   | 0   | 0   | 12  |
| CO | 0   | 217 | 7   | 0   | 1   | 0   | 0   | 0   | 1   |
| DW | 0   | 0   | 215 | 3   | 0   | 1   | 0   | 0   | 1   |
| EA | 2   | 0   | 2   | 200 | 8   | 0   | 0   | 0   | 4   |
| OT | 35  | 3   | 0   | 6   | 156 | 1   | 0   | 0   | 11  |
| SA | 0   | 0   | 2   | 1   | 3   | 220 | 0   | 0   | 0   |
| VC | 0   | 0   | 0   | 0   | 0   | 0   | 222 | 0   | 0   |
| WT | 0   | 0   | 0   | 0   | 0   | 0   | 0   | 214 | 0   |
| WO | 17  | 1   | 2   | 4   | 15  | 0   | 0   | 0   | 173 |

Besides the confusion matrix of combined system, the comparison of the baseline system and the proposed system in terms of F1-score is shown as in Table VS. It can be observed the performances of all other eight activities are improved expect that of absence. Meanwhile, the F1-scores of dishwashing, eating, and other are remarkably improved by 19.25%, 9.38%, and 32.88% respectively. Therefore, the macro-averaged F1-score is improved significantly by 7.69%, achieving 92.19%.

Table V Comparison of the baseline system and the proposed system

| Activity | F1-score of baseline system | F1-score of proposed system |
|----------|------------------------------|------------------------------|
| Absence | 85.41% | 84.45% |
| Cooking | 95.14% | 97.09% |
| Dishwashing | 76.73% | 95.98% |
| Eating | 83.64% | 93.02% |
| Other | 44.76% | 77.61% |
| Social activity | 93.92% | 97.99% |
| Vacuum cleaning | 99.31% | 100% |
| Watching TV | 99.59% | 100% |
| Working | 82.03% | 83.57% |
| Macro-averaged F1-score | 84.50% | 92.19% |

## 5. CONCLUSION

In this paper, we effectively address Task 5 by proposing a neural network based system, consists of 2D CNN, 1D CNN, and stateless LSTM. Power spectrogram feature and MFCCs feature is extracted after pre-processing the raw data. By applying the proposed system to the development dataset, high performance can be achieved as the macro-averaged F1-score is improved from 84.50% to 92.19%, compared with the baseline system provided by the DCASE committee.

## 6. REFERENCES

[1] F. Erden, S. Velipasalar, A. Z. Alkar, and A. E. Cetin, "Sensors in Assisted Living: A survey of signal and image processing methods," *IEEE Signal Processing Magazine*, vol. 33, no. 2, pp. 36-44, Mar.2016.

[2] M. Yu, A. Rhuma, S. M. Naqvi, L. Wang, and J. Chambers, "A posture recognition-based fall detection system for monitoring an elderly person in a smart home environment," *IEEE Trans. Inf. Technol. Biomed.*, vol. 16, no. 6, pp. 1274-1286, 2012.

[3] N. K. Suryadevara, S. C. Mukhopadhyay, "Determining wellness through an ambient assisted living environment," *IEEE Intell. Syst.*, vol. 29, no. 3, pp. 30-37, 2014.

[4] E. Principi, P. Olivetti, S. Squartini, R. Bonfigli, F. Piazza, "A floor acoustic sensor for fall classification," in *Proc. 138th Audio Engineering Soc. Convention*, pp. 949-958, 2015.

[5] G. A. Koshmak, M. Linden, A. Loutfi, "Evaluation of the android-based fall detection system with physiological data monitoring," in *Proc. Annu. Int. Conf. IEEE Engineering in Medicine and Biology Society*, pp. 1164-1168, 2013.

[6] K. Ozcan, A. K. Mahabalagiri, M. Casares, and S. Velipasalar, "Automatic fall detection and activity classification by a wearable embedded smart camera," *IEEE J. Emerg. Sel. Top. Circuits Syst.*, vol. 3, no. 2, pp. 125-136, 2013.

[7] A. K. Bourke, S. Prescher, F. Koehler, V. Cionca, C. Tavares, S. Gomis, V. Garcia, and I. Nelson, "Embedded fall and activity monitoring for a wearable ambient assisted living solution for older adults," in *Proc. Annu. Int. Conf. IEEE Engineering in Medicine and Biology Society*, pp. 248-251, 2012.

[8] DCASE 2018 challenge. [Online]. Available: http://dcase.community/challenge2018/

[9] SINS. Sound Interfacing through the Swarm. [Online]. Available: http://www.esat.kuleuven.be/sins/

[10] Fonseca, Eduardo et al, "General-purpose Tagging of Freesound Audio with AudioSet Labels: Task Description, Dataset, and Baseline,".

[11] Graves, Alex and Jaitly, Navdeep and Mohamed, Abdel-rahman, "Hybrid speech recognition with deep bidirectional LSTM," in *Proc. Automatic Speech Recognition and Understanding (ASRU), 2013 IEEE Workshop on*, pp. 273-278, 2013.

[12] Fukushima, Kunihiko and Miyake, Sei, "Neocognitron: A self-organizing neural network model for a mechanism of visual pattern recognition," *Competition and cooperation in neural nets*.

[13] LeCun, Yann et al, "Gradient-based learning applied to document recognition," *Proceedings of the IEEE.*, vol. 86, no. 11, pp. 2278--2324, 1998.

[14] Prasoon, Adhish et al, "Deep feature learning for knee cartilage segmentation using a triplanar convolutional neural

network," *International conference on medical image computing and computer-assisted intervention.*, pp. 246-253, 2013.

[15] A. Graves, N. Jaitly, and A. Mohamed, "Hybrid speech recognition with Deep Bidirectional LSTM, " in *2013 IEEE Workshop on Automatic Speech Recognition and Understanding*, Olomouc, 2013, pp. 273-278.